# A spin light emitting diode incorporating ability of electrical helicity switching

N. Nishizawa[a], K. Nishibayashi, and H. Munekata

Imaging Science and Engineering Laboratory, Tokyo Institute of Technology

4259-J3-15 Nagatsuta, Midori-ku, Yokohama 226-8503, Japan

**ABSTRACT**

Fabrication and optical characteristics of a spin light-emitting-diode (spin-LED) having dual spin-injection electrodes with anti-parallel magnetization configuration are reported. Alternating a current between the two electrodes using a PC-driven current source has led us to the observation of helicity switching of circular polarization at the frequency of 1 kHz. Neither external magnetic fields nor optical delay modulators were used. Sending dc-currents to both electrodes with appropriate ratio has resulted in continuous variation of circular polarization between the two opposite helicity, including the null polarization. These results suggest that the tested spin-LED has the feasibility of a monolithic light source whose circular polarization can be switched or continuously tuned all electrically.

---

[a] Electronic mail: nishizawa@isl.titech.ac.jp

As represented by the Poincaré sphere, light can take two polarization states, the linear polarization and the circular polarization, and superposition between the two states. The former is the state with which we deal occasionally in various applications in optics and photonics. The latter, because it has a spin angular momentum, is the state which works as a unique optical tool to investigate chirality of molecules, crystals, and spin structures, as well exemplified by the chiral resolution technique used in synthetic chemistry[1]. There have been other applications proposed on the basis of the difference in the optical response of chiral structures as well as quantum orthogonality between two opposite helicities of circularly polarized light: examples are all-optical magnetic writing[2], three-dimensional display[3], and quantum optical communications[4]. However, the lack of a monolithic light source with ability of electrical helicity switching impedes advancing the development of those applications. In this letter, we propose and demonstrate a prototype, circularly polarized light source having the ability of electrical helicity switching.

A light emitting diode combined with a magnetic, spin-injection electrode (SIE), nowadays called a spin light-emitting-diode (spin-LED), is a light source that inherently emits circularly polarized light through the radiative recombination between spin-polarized electrons and holes[5, 6]. It has been used in laboratories as the instrument to optically quantify spin polarization of injected carriers in a semiconductor[7-10]. For practical applications of spin-LED, operation at room temperature, coherent emission from laser structures, and development of means for electrical helicity switching are desired. Among those, the former two subjects have been initiated already by other workers[11-16]. The third subject, however, has never been studied up to now. The present work, although our experimental demonstration might be the egg of Columbus, aims at showing the simplest but the most practical means of electrical helicity switching by fabricating on a top surface of the LED a pair of SIEs whose remnant magnetization is anti-parallel to each other, and demonstrating two functionalities associated with direct control of circular polarization: electrical helicity switching and electrical polarization blending.

Tested spin-LED chips consist of AlGaAs/InGaAs/AlGaAs double-heterostructure (DH) and two Fe-based SIEs formed side-by-side with anti-parallel magnetization configuration on top of the DH. We call hereafter a pair of those SIEs the dual-SIE. Alternately sending a current to the two SIEs has resulted in electrical helicity switching between the circular polarization of $P_{EL} \sim +0.1$ or $-0.1$ with the alternation frequency of 1 kHz set by a PC-driven current source. Furthermore, sending currents synchronously to the dual-SIE with the appropriate current ratio has resulted in continuous change in circular polarization. These results demonstrate the feasibility of a monolithic, single-point light source whose circular polarization can be switched or tuned all electrically without using external magnetic fields or peripheral optical delay modulators, such as piezo-optic and magneto-optic modulators.

The actual device consists of, from the top, two polycrystalline Au (20 nm) / Ti (5 nm)

/ Fe multi-layers with Fe layer thicknesses of 100- and 30-nm, a crystalline AlO$_x$ tunnel barrier layer (1 nm), and an AlGaAs/InGaAs DH grown on a *p*-type GaAs (001) substrate. The DH and the AlO$_x$ layer both were prepared by using a molecular beam epitaxy system. The DH consisting of 300-nm *n*-Al$_{0.1}$Ga$_{0.9}$As (Sn ~ 1×10$^{17}$ cm$^{-3}$) / 15-nm undoped Al$_{0.1}$Ga$_{0.9}$As / 500-nm undoped In$_{0.03}$Ga$_{0.97}$As / 20-nm undoped Al$_{0.2}$Ga$_{0.8}$As / 500-nm *p*-Al$_{0.2}$Ga$_{0.8}$As (Be ~ 1×10$^{18}$ cm$^{-3}$) / 500-nm *p*-GaAs (Be ~ 1×10$^{18}$ cm$^{-3}$) / *p*-GaAs (001), was prepared at the substrate temperature of $T_s$ = 510 ºC, whereas the crystalline AlO$_x$ (x-AlO$_x$) tunnel barrier was formed with great care at the room temperature[17]. The sample wafer was then cleaved into numbers of 20-mm long and 5-mm wide slabs. The slabs were loaded in the electron-beam evaporation chamber, in which pairs of Au/Ti/Fe SIE stripes with two different Fe-layer thicknesses of 100 nm and 30 nm were deposited with spatial separation of 0.3 mm using mechanical masks. Long sides of the slabs and the stripes were both parallel to the GaAs <1$\bar{1}$0> axis. We call hereafter the 100-nm and 30-nm Fe SIEs as SIE-100 and SIE-30, respectively. After annealing the slabs at 230 ºC for 1 hour in a nitrogen gas flow, the backside of the *p*-GaAs substrate was mechanical polished to the thickness of around 200 μm. Finally, slabs were cleaved into numbers of 3 mm × 2 mm dual-SIE spin-LED chips having two, 1.3 mm × 2.0 mm, rectangular SIEs side-by-side on the top surface.

Due to the shape anisotropy, the magnetization easy axis is parallel to the long side of the rectangular SIE which is along the <1$\bar{1}$0> axis. Shown in Fig. 1(a) are normalized magnetization hysteresis curves ($M$–$H$ curves) of SIE-100 (blue lines), SIE-30 (green lines), and dual-SIE (red lines). Measurements were carried out at 5 K by a magnetometer equipped with a superconducting quantum interference device (SQUID) with magnetic fields applied along the in-plane <1$\bar{1}$0> axis. Magnetization of the SIE-100 is reversed at $H \approx \pm 90$ Oe, whereas that of the SIE-30 at $H \approx \pm 220$ Oe. The observed difference in the switching field is explained qualitatively in terms of the difference in the energy of reverse domain which is known to increase with decreasing the layer thickness[18]. The $M$–$H$ curve of the dual-SIE is well reproduced by the superposition of the two $M$–$H$ curves of SIE-100 and SIE-30, indicating no significant degradation during device processing.

The dual-SIE spin-LED chips were placed on a Cu chip carrier, and Cu micro-probes, which were pre-assembled in the chip carrier, were pressed on the surfaces of dual-SIE to establish electrical contacts (Fig.1 (b)). No silver paste was used. The chip carrier was then loaded into a GM-refrigeration optical cryostat with the cleaved (1$\bar{1}$0) facet of the chips facing a cryostat window, and was cooled down to 5 K. As shown also in Fig.1(b), circular polarization of electroluminescence (EL) from the (1$\bar{1}$0) facet, $P_{EL}(h\nu)=\{I(\sigma^+)-I(\sigma^-)\}/\{I(\sigma^+)-I(\sigma^-)\}$, was measured by the double lock-in technique using an optical chopper operated at 200 Hz and a photo-elastic modulator (PEM) operated at 50 kHz, unless specified otherwise. Here, $I(\sigma^+)$ and $I(\sigma^-)$ are, respectively, the EL intensity of $\sigma^+$ and $\sigma^-$ components. In our experiments, spin polarization of injected electrons that recombine in a thick active $In_{0.03}Ga_{0.97}As$ layer is represented by the relation $P_{spin} = 2 \times P_{EL}$[7,8]. Anti-parallel magnetization configuration in the dual-SIE was achieved by first applying the external field of $H = +5$ kOe along the $<1\bar{1}0>$ axis to realize parallel configuration, then applying $H = -180$ Oe to only reverse magnetization of SIE-100, and finally removing the external field. Shown in Fig. 1(c) is the far-field EL emission pattern from the tested spin-LED chip with currents sent synchronously to both SIEs. Observation of two EL emission points just below the two Cu microprobes has suggested that electrons injected from SIEs do not diffuse laterally at low temperatures. Those two emission points were focused by using the lens with the magnification ratio of 2.5 on a single, wide-window optical fiber detector (6-mm width), and hence could be practically regarded as a single-point light source. Improvement towards the single-point light source will be discussed in the later paragraphs.

Figure 2 (a) and (b) show EL spectra, EL-100 and EL-30, obtained by sending a current to SIE-100 and SIE-30, respectively. Current density was 2.5 A/cm$^2$ for both cases. The shapes of both spectra are almost identical with the EL peak position at 1.48 eV which matches well with the band gap energy of the active In$_{0.03}$Ga$_{0.97}$As layer. The full width at half maximum is around 40 meV. Reflecting the anti-parallel configuration, the values of circular polarization (carrier spin polarization) are $P_{EL-100} \sim +0.12$ ($P_{SIE-100} \sim +0.24$) for EL-100, whereas $P_{EL-30} \sim -0.09$ ($P_{SIE-30} \sim -0.18$) for EL-30. Estimation of both $P_{EL}$ values were carried out by integrating $P_{EL}(h\nu)$ over the photon energy region of $h\nu = 1.45 - 1.52$ eV. The highest $P_{EL}$ values obtained in the present work was $|P_{EL}| = 0.15$. With this value, the initial spin polarization of injected carriers is inferred to be $|P_{spin(initial)}| = 0.39$ referring to the spin diffusion length of around 1 μm [19]. Fringe-field-induced magneto-optical effects in the InGaAs active and AlGaAs adjacent layers are not responsible for the observed circular polarization because of the following two reasons: firstly, as clarified through the study with lateral-type MnSb/GaAs spin-LED[20, 21], the field strength in the 300-nm deep InGaAs active layer is 150 Oe or lower, and secondly, the values of *g*-factor for the conduction band in the AlGaAs clad layers are nearly zero[22, 23].

Experiments of electrically helicity switching was carried out by sending a rectangular-shape ac current of 1 kHz to each SIE with a phase shift of a half cycle (Fig.1(b)). The current density was 2.5 A/cm$^2$ for both SIEs. The data shown in Fig.2 (c) were $P_{EL}$ (1.48 eV) values taken with the double lock-in technique (Fig.1 (b)), whereas those in Fig.2 (d) were taken by substituting a quarter-wave plate for a piezo-optic modulator. In both measurements, abrupt switching of the optical helicity is clearly observed. This fact indicates that helicity switching is achieved by the proposed dual-SIE spin-LED without using any peripheral optical delay modulators. Neglecting the LRC time constant of the spin-LED and circuit, we infer that switching frequency in the range of GHz is possible, since the speed of helicity switching is limited by the radiative recombination time which is sub-ns to ns at cryogenic temperatures[24]. Pulsed laser emission with alternating circular polarization synchronized to the electron Larmor precession at 22 GHz was demonstrated by circularly polarized optical excitation of a semiconductor microcavity in a transverse magnetic field [12].

Shown in Fig. 3 are temperature dependences of normalized EL intensity, $I_{EL}$, and normalized circular polarization, $P_{EL}$, for both EL-100 and EL-30. We first anticipated that the temperature dependences would be nearly same between the two EL emissions, reflecting homogeneous electronic and optical qualities of DH. They, however, appeared to be different. For EL-100, the $I_{EL}$ value decreases gradually up to around 50 K beyond which it decreases relatively steeper. The reduced $I_{EL}$ at $T > 50$ K is attributed to the non-radiative recombination via deep levels around the middle of the band gap in the InGaAs active layer, as confirmed by the model calculation (a dashed line in Fig.3) based on the bimolecular rate equation[25] and the Shockley-Read recombination model[26]. On the other hand, the $P_{EL}$ value follows $P_{EL} \propto T^{-0.2}$ at $T < 30$ K and $P_{EL} \propto T^{-2.5}$ at $T \geq 30$ K (inset Fig.3), which is reminiscent of the Elliott-Yafet (EY) and D'yakonov-Perel' (DP) mechanisms in GaAs at low and high temperature regions, respectively[27]. Turning eyes on EL-30, both $I_{EL}$ and $P_{EL}$ values decrease abruptly with almost same temperature dependence from the initial temperature of $T = 5$ K. Taking account of relatively weaker emission intensity (Fig. 1(c)) and smaller $P_{EL}$ value of the EL-30, the observed steep reductions in $I_{EL}$ and $P_{EL}$ with the SIE-30 would be attributed to some relaxation path which belongs with Fe/x-AlO$_x$/n-AlGaAs junction of SIE-30 and affects both $I_{EL}$ and $P_{EL}$. We leave this problem for the future study. We are also able to point out, from the observed temperature dependence, that MO effect in SIEs is not the primary mechanism for the polarized

EL.

Within the range of a current density studied in this work (< 3.0 A/cm$^2$), $P_{EL}$ values do not degrade significantly with increasing a spin-injection current, as shown in the two insets in Fig.4. On the basis of those data, the emission intensities of EL-100 and EL-30, $I_{100}$ and $I_{30}$, respectively, were tuned synchronously by sending different amount of currents to SIE-100 and SIE-30 to test whether the circular polarization of EL can be controlled electrically. Results of experiments are shown in Fig. 4. Here, the horizontal axis represents the ratio of emission intensities, $I_{100} / (I_{30} + I_{100})$, and the vertical axis the detected circular polarization. In this experiment, the current sent to each SIE was adjusted in such a way to keep the total EL intensity ($I_{30} + I_{100}$) constant. It is clearly seen in the data that circular polarization can be controlled continuously between negative and positive helicities by tuning the currents sent to two SIEs. The null circular polarization at the intensity ratio of 0.4 did not contain any linear polarization, as was verified by the separate experiments using a linear polarizer. This fact indicates that two EL emissions occur spontaneously without any mechanism which correlates the coherency between the two emissions. The data shown in Figs.2 and 4 suggest that two functionalities can be realized depending on the operation mode of the dual-SIE spin-LED: the alternate SIE operation gives rise to electrical polarization switching, whereas the synchronous SIE operation does electrical polarization blending.

For the rest of this Letter, we address problems which need further studies in view of a light source. The first problem is of the observed double emission points (Fig.1 (c)). This problem is due to the limited diffusion length of injection carriers in semiconductor layers, and can be solved by engineering the resistance distribution in the LED structure: formation of a current confinement region in transport and/or active layers is one of the most practical approaches[28, 29]. Another approach is introducing means of electrically switching magnetization vector in the single SIE. Imagine placing on top of an LED a single ferromagnetic layer incorporating a pair of 180-deg. domains with a pair of electrical contacts touching each domain. The domain wall moves back and forth by sending an ac-current across the ferromagnetic layer[30], by which spin polarization of carriers injected in the LED can be reversed. Helicity switching frequency of 1 MHz would be achieved when the effective length of SIE and the velocity of domain wall propagation are 25 μm and 50 m/s at room temperature[31], respectively. Spin filtering using resonant tunneling[32, 33] or magnetic insulators layers[34] would be another interesting option presupposing that spin states are selected by electrical means.

The second problem concerns with fundamental limitation in the maximum circular polarization of $P_{EL} = 0.5$ which is due to the use of a bulk active layer[7] in order to accommodate in-plane spin-polarized carriers injected from the SIE of an in-plane ferromagnet. We suppose that this limitation could be overcome by the mechanism of light amplification by stimulated emission radiation, since the optical transition of one particular helicity would benefit an optical gain[15]. This mechanism might also compensate the disadvantage of spin relaxation at high temperatures. Other approach to circumvent the problem of $P_{EL} = 0.5$ is to utilize a quantum-well active layer together with out-of-plane SIEs. In this case, recombination between the first sub-band electrons and heavy holes is the dominant optical transition and thus $P_{EL} = 1.0$ for the light emitted vertically from the surface. Fabrication of the dual-SIEs with out-of-plane, anti-parallel magnetization configuration would be the key subject for this approach.

In summary, GaAs-based spin-LED chips incorporating a pair of Fe-based spin injection electrodes with anti-parallel magnetization configuration (dual-SIE spin-LEDs) have been prepared and tested in view of a light source having the ability of electrical helicity switching of circular polarization. Alternately sending a current to two different Fe electrodes at the frequency of 1 kHz has resulted in helicity switching without using an external magnetic field or peripheral optical delay modulators. Sending currents synchronously to the two SIEs with appropriate current ratio has resulted in continuous change between two opposite circular polarization, including the null circular polarization. On the basis of these observations, we have concluded that the tested device would have a feasibility of a monolithic circularly-polarized light source with two functionalities: all-electrical helicity switching and all-electrical polarization blending.

Authors are grateful to T. Takita and N. Hieda for their technical assistance in various experiments, and T. Matsuda for theoretical discussions. We acknowledge partial supports from Advanced Photon Science Alliance Project from MEXT and Grant-in-Aid for Scientific

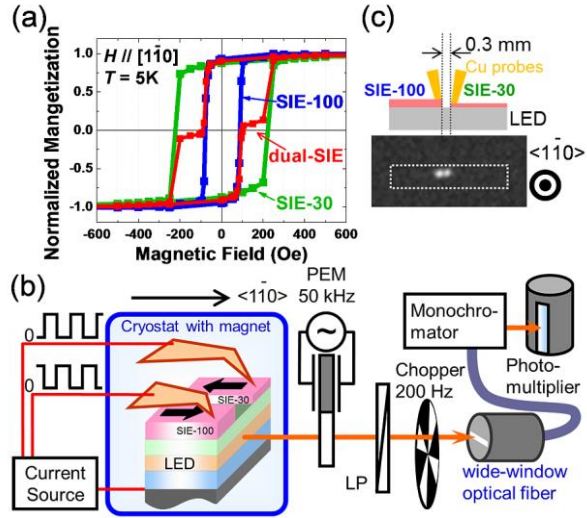

FIG. 1. (Color online) (a) Normalized *M-H* curves for SIE-100 (blue), SIE-30 (green), and dual-SIE on a LED chip (red). (b) Schematic illustration of experimental setups. Electroluminescence (EL) from a cleaved facet was guided into a monochrometer through a wide-window optical fiber. Lens is omitted from an illustration. (c) Far-field EL pattern observed from one of the tested chips. Currents were sent synchronously to both SIEs with current density of $J \sim 3.2$ A/cm$^2$.

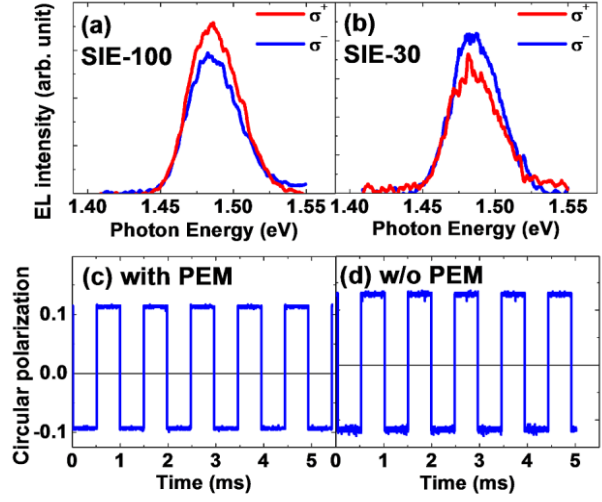

FIG. 2. (Color online) EL spectra, EL-100 and EL-30, obtained at 5 K by sending currents ($J \sim 2.5$ A/cm$^2$) to (a) SIE-100 and (b) SIE-30, respectively. (c) Experimental data which demonstrate 1-kHz helicity switching obtained by using experimental setups shown in Fig.1(b), and (d) those obtained without using a PEM.

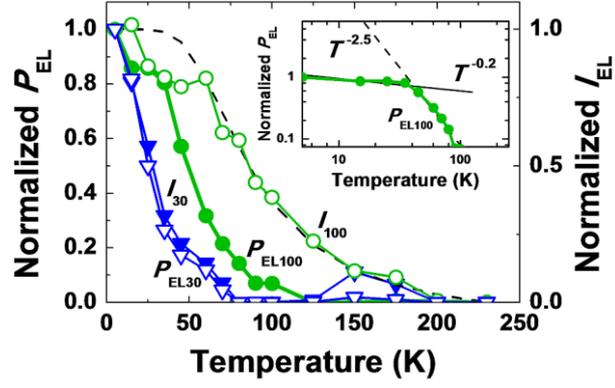

FIG. 3. (Color online) (a) Temperature dependence of absolute $P_{EL}$ values normalized by the value at 5 K and EL intensity $I_{EL}$ normalized by the value at 5 K, for EL-100 (closed and open circles) and EL-30 (closed and open triangles), respectively. Lines between plots are for eye guides. A dashed line represents the calculated internal quantum efficiency $\eta$. The lifetime of minority carrier and the energy level of the non-radiative states from the conduction band edge are assumed to be $\tau_{p0} = 0.5$ ns and $E_{\text{non-rad}} = 0.77$ eV, respectively, for model calculation. Inset shows logarithmic plots of $P_{EL}$ vs. $T$ for EL-100 with fitting lines in the form of $P_{EL} = T^{-n}$.

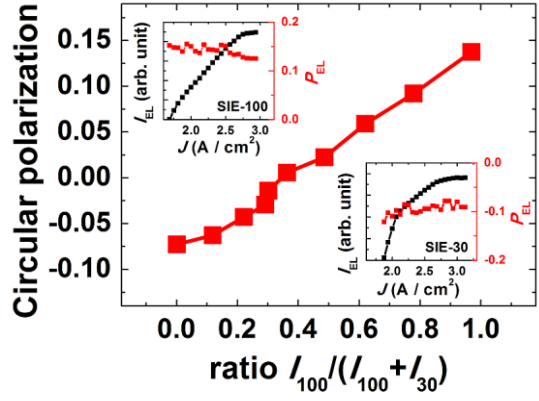

FIG. 4. (Color online) A plot of circular polarization vs. ratio of EL emission intensity $\{I_{100} / (I_{100} + I_{30})\}$. Inset figures show the emission intensity $I$ (black dots) and $P_{EL}$ (red dots) as a function of a forward current density for EL-100 (upper left) and EL-30 (lower right). $\Delta P_{EL} = \pm 0.03$ in the region of 1.7~ 2.5 A/cm$^2$ for both emissions.